# The mechanisms behind extreme susceptibility of photon avalanche emission to quenching

M.Majak*, M.Misiak, A.Bednarkiewicz*

Institute of Low Temperature and Structure Research, Polish Academy of Sciences, ul. Okólna 2, 50-422 Wroclaw, Poland

Corresponding authors: m.majak@intibs.pl, a.bednarkiewicz@intibs.pl

**Abstract**

The photon avalanche (PA) process that emerges in lanthanide-doped crystals yields a threshold and highly nonlinear (of the power law order > 5) optical response to photoexcitation. PA emission is the outcome of excited-state absorption combined with a cross-relaxation process, which creates positive and efficient energy looping. In consequence, this combination of processes should be highly susceptible to small perturbations in energy distribution and thus can be hindered by other competitive "parasite" processes such as energy transfer (ET) to quenching sites. Although luminescence quenching is a well-known phenomenon, the exact mechanisms of susceptibility of PA to resonant energy transfer (RET) remain poorly understood limiting practical applications. A deeper understanding of these mechanisms may pave the way to new areas of PA exploitation. This study focuses on the investigation of the LiYF$_4$:3%Tm$^{3+}$ PA system co-doped with Nd$^{3+}$ acceptor ions, which was found to impact both the looping and emitting levels and thus to effectively disrupt PA emission, causing an increase in the PA threshold ($I_{th}$) and a decrease in PA nonlinearity ($S_{max}$). Our complementary modelling results revealed that the ET from the looping level increased $I_{th}$ and $S_{max}$, whereas the ET from the emitting level diminished $S_{max}$ and the final emission intensity. Ultimately, significant PA emission quenching demonstrates a high relative sensitivity ($S_R$) to infinitesimal amounts of Nd$^{3+}$ acceptors, highlighting the potential for PA to be utilized as an ultra-sensitive, fluorescence-based reporting mechanism that is suitable for the detection and quantification of physical and biological phenomena or reactions.

## 1. Introduction

A photon avalanche (PA) is a special type of energy transfer up-conversion that yields a highly nonlinear pump-power-dependent luminescence emission intensity $I_L = (I_P)^S$,[1] where the order of nonlinearity S ranges from 5 to 46.[2–4] In comparison, an ordinary up-conversion process generally achieves an S equal 2–4.[5,6] High nonlinearity grants the possibility of deploying PA materials in many applications, such as sensing of physical, chemical, and biological quantities,[7–10] sub-diffraction imaging,[2,3,11–14] reservoir computing,[15] and data storage,[16] and may become a key enabling technology for security inks and solar energy harvesting.[17] The PA process can only occur after the fulfillment of two primary conditions (**Figure 1a**)[1,18,19]. First of all, the ground-state absorption (GSA) must be negligible with regard to the excited-state absorption (ESA), and thus, the ratio $\sigma_{ESA}/\sigma_{GSA}$ of absorption cross sections must be equal or higher than typically 10$^4$.[17,20] After the initial multiphonon-assisted (MPR) GSA from the ground (GL) to the looping level (LL), an efficient ESA occurs and populates the emitting level (EL). From this level, a few processes can occur, including other ESA to higher states,[21,16] radiative and non-radiative recombinations, non-radiative relaxations, or cross-relaxation (CR) energy transfer. The emergence of a CR between two adjacent lanthanide ions is the second critical requirement for efficient energy looping and storage. During CR, the population of the looping level doubles with each iteration of the looping cycle. After achieving a certain population of the looping state across many lanthanide ions, which is realized by increasing the excitation power up to a threshold level, the material becomes a good absorber (through ESA), and the luminescence emission intensity starts to grow in a highly nonlinear manner with a further increase in the photoexcitation intensity.

PA is a highly nonlinear phenomenon governed by complex energy distribution dynamics and many interdependent processes and can be easily perturbed by numerous quenching processes such as unintentionally co-doped impurities, crystal vacancies, dislocations, and other crystal defects.[22–24] Additionally, in nanomaterials, surface quenching by ligands and solvent molecules becomes an additional path for disrupting PA energy looping and emission.[24] Initial studies on avalanching nanoparticles (ANPs) with varying shell thicknesses demonstrated that any process capable of modifying the relaxation rate of the looping state (e.g., surface quenching) affects $I_{th}$.[24] Nevertheless, these negative processes, when properly managed and understood, may become novel, sensitive luminescence reporters for the (bio)sensing of chemical surroundings. For example, it has been suggested that the perturbation of PA emission in nanomaterials by resonant surface acceptors may effectively increase the resonant energy transfer (RET) sensitivity range beyond the Förster distance.[25] Nevertheless, the specific contributions of the energy transfer (ET) from the looping and emitting states are still not entirely understood, and without experimental evidence remain speculative. While conventional understanding of luminescence quenching refers to quenching of the emitting levels, PA is significantly more complex as it is collective and excitation history dependent phenomenon. The major motivation and novelty of the studies presented here is to both qualitatively and quantitatively understand the principles behind PA quenching, as well as critically and generically discuss these phenomena with future applications of PA emission in mind.

Because every PA system involves at least three states (**Figure 1a**), ground ($n_1$), looping ($n_2$), and emitting ($n_3$) levels, we distinguished three scenarios of PA quenching by the resonant acceptor (**Figure 1b**): ($Q_I$) ET from the looping level, ($Q_{II}$) ET from the PA-emitting level, and ($Q_{III}$) a combination of both. The latter reflects the real-world scenario studied here, where the acceptor possesses a complex energy-level structure with many absorption bands. In the presence of such a $Q_{III}$ acceptor, an increased pump power threshold for PA emission is expected (**Figure 1c**) because the energy transfer from the looping and emitting states inhibits the population growth of the looping state. Furthermore, it is anticipated that the energy transfer from both levels will reduce the emission intensity of PA. The order of nonlinearity in the PA emissions is also expected to decrease as the CR competes with the new non-radiative process. Compared with pure PA (**Figure 1d**, PA left branch), the presence of energy acceptors should hinder the excited-state population buildup and photon emission (**Figure 1d**, PA+RET right branch) by disturbing the smooth and efficient ESA ($Q_I$) and CR ($Q_{II}$), deferring energy looping directly ($Q_I$) or indirectly ($Q_{II}$).

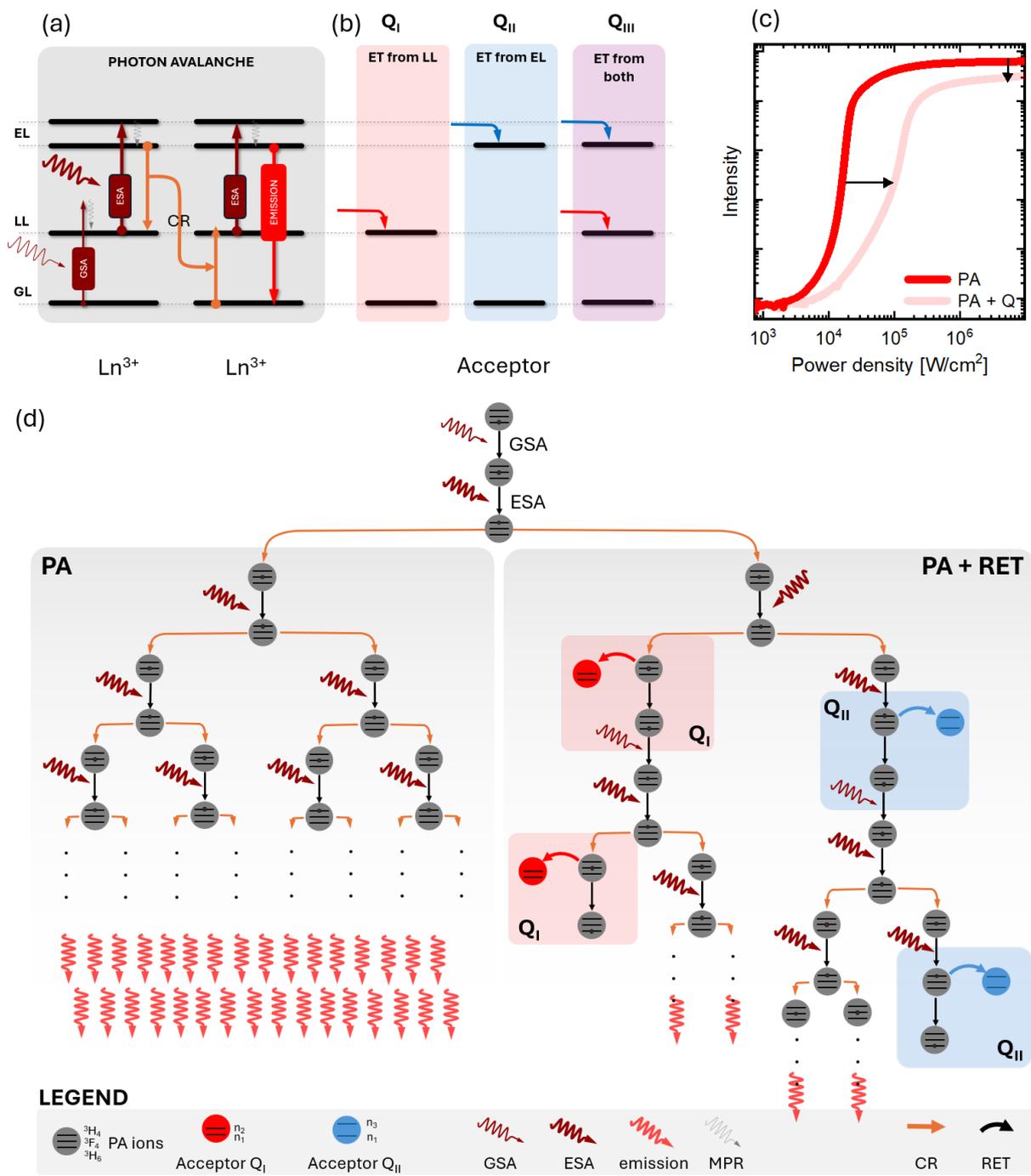

*Figure 1 Explanation of PA quenching. (a) Simpliefied energy-level diagram showing PA process: inefficient GSA from ground level (GL) is followed by efficient ESA, energy CR in a positive loop, followed by photon emission; (b) diagram illustratrating the acceptor energy levels in close proximity to the looping and emitting levels of lanthanide ions, demonstrating three distinct scenarios of resonant energy transfer (ET) pathways leading to quenching of the PA emission: (I) D-A ET from the looping level LL ($Q_I$), (II) D-A ET from the emitting level EL ($Q_{II}$), and (III) D-A ET from both looping and emitting levels of D emitting ion ($Q_{III}$); (c) simulated "S" shape data schematically showing the impact of quenching on the PA emission characteristics; (d) diagram schematically compares pure PA with acceptor PA+RET and highlighs the*

*mechanisms of RET quenching within the whole branches of energy looping ions subnetwork leading to less efficient PA emission and increased PA treshold.*

In our study, we take advantage of the remarkable PA emission exhibited by $Tm^{3+}$ ions. The $Tm^{3+}$ ions were incorporated into a crystalline matrix of $LiYF_4$, with a 3% doping level to assure PA emission generation and simultaneously avoid potential photodarkening effect (explained by PA electron trapping in surface defects) that has been observed in nanocrystals doped with above 8% $Tm^{3+}$ ions [26]. This concentration not only shows satisfactory nonlinear properties to understand the general principles behind PA quenching, but also attains a lower power-excitation threshold compared with higher concentrations.[2] PA in $Tm^{3+}$ ions can be achieved by non-resonant GSA at 1059 nm, leading to the excitation of $^3H_5$ level (**Figure 2a**). Non-radiative relaxation $^3H_5 \rightarrow {}^3F_4$ populates the long-living looping state. As soon as the looping state becomes preliminarily populated, the resonant ESA process further excites higher-lying energy levels such as $^3F_3$ and $^3F_2$. With another non-radiative relaxation $^3F_3/^3F_2 \rightarrow {}^3H_4$ the emitting state becomes populated. Both GSA and ESA must occur with phonon assistance to populate the crucial looping and emitting states. With populated $^3H_4$ level, efficient CR occurs between adjacent ions, doubling the $^3F_4$ population and further enhancing the ESA until saturation (SAT) is reached. Radiative recombination at the $^3H_4$ level yields the most intense PA emission at 800 nm (**Figure 2b**).

To understand the quenching of the two key levels on the PA excitation and emission, the capability of individual quenching of the emitting and the looping levels are critical. Although a wide range of organic or inorganic species exhibit absorption bands corresponding to the $Tm^{3+}$ $^3H_4$ emitting level (800 nm) (e.g., IR808, $Ag_2S$ QDs, etc.), there is a limited availability of those that can resonantly and specifically interact with the looping state ($^3F_4$ level E~6000 $cm^{-1}$, ca. at ~1670 nm). Even if such two acceptors to be anchored on the surface of the ANP are available, their conjugation chemistry and ANP surface coverage may not be homogenous, thus, such experiment would not allow to resolve the question about the nature, mechanism and importance of the two levels in PA emission quenching. To make the studies reliable and comparable, we evaluated how the intended co-doping of the $Tm^{3+}$ energy-looping system with various concentrations of $Nd^{3+}$ acceptor ions influenced the processes leading to PA emission from the $Tm^{3+}$ ions. The other lanthanides (e.g. $Er^{3+}$, $Ho^{3+}$ etc.) could help to elucidate the mechanism of quenching of PA emission, but we have deliberately chosen $Nd^{3+}$ ions, whose energy levels overlap with the two critical levels in $Tm^{3+}$ that are involved in PA machinery. Moreover, up to 1% $Nd^{3+}$ doping, $Nd^{3+}$ does not show significant concentration quenching in fluoride hosts [27]. In particular, neodymium possess energy levels $^4F_{5/2}$ and $^2H_{9/2}$ (12005-12933 $cm^{-1}$), which energetically overlap with the emitting $Tm^{3+}$ level $^3H_4$ (12180 to 12891 $cm^{-1}$) as well as $^4I_{15/2}$ (5328-6432 $cm^{-1}$) level energetically overlapping with the $Tm^{3+}$ looping level $^3F_4$ (5180-5972 $cm^{-1}$).[28] In this case, the PA quenching should arise from both the emitting and looping levels, which corresponds to the $Q_{III}$ quenching scheme. Consequently, it is challenging to split and discuss the specific effects of quenching resulting from separate ET processes originating from either the emitting or looping level. However, to investigate the individual contributions to PA quenching, we employed photon avalanche differential rate equations (PADRE) to model the behavior of PA luminescence. Furthermore, considering the many nonspecific quenching mechanisms that could potentially influence PA emission in nanoparticles [24], coupled with the potential related to NP passivating shell formation (i.e. shell thickness variability, dopant concentration gradient across the NPs, local and particle-to-particle inhomogeneities, ion intermixing / migration between subsequent NP layers, access of surface ligands and solvent vibration, surface defects) we deliberately mitigated these concerns. This was achieved by utilizing large microcrystals (ca. 60 μm) (**Figure 2c, S1**) and investigating PA properties from deep within the subsurface volume of such microcrystals using a tightly focused laser beam (**Figure d,** $\phi$=1.2 μm diameter, i.e. ca. 50-fold smaller than the single microcrystal). The structural purity and uniform morphology of these samples were confirmed by X-ray diffraction (XRD) and scanning electron microscopy (SEM) (**Figure S1, S2**).

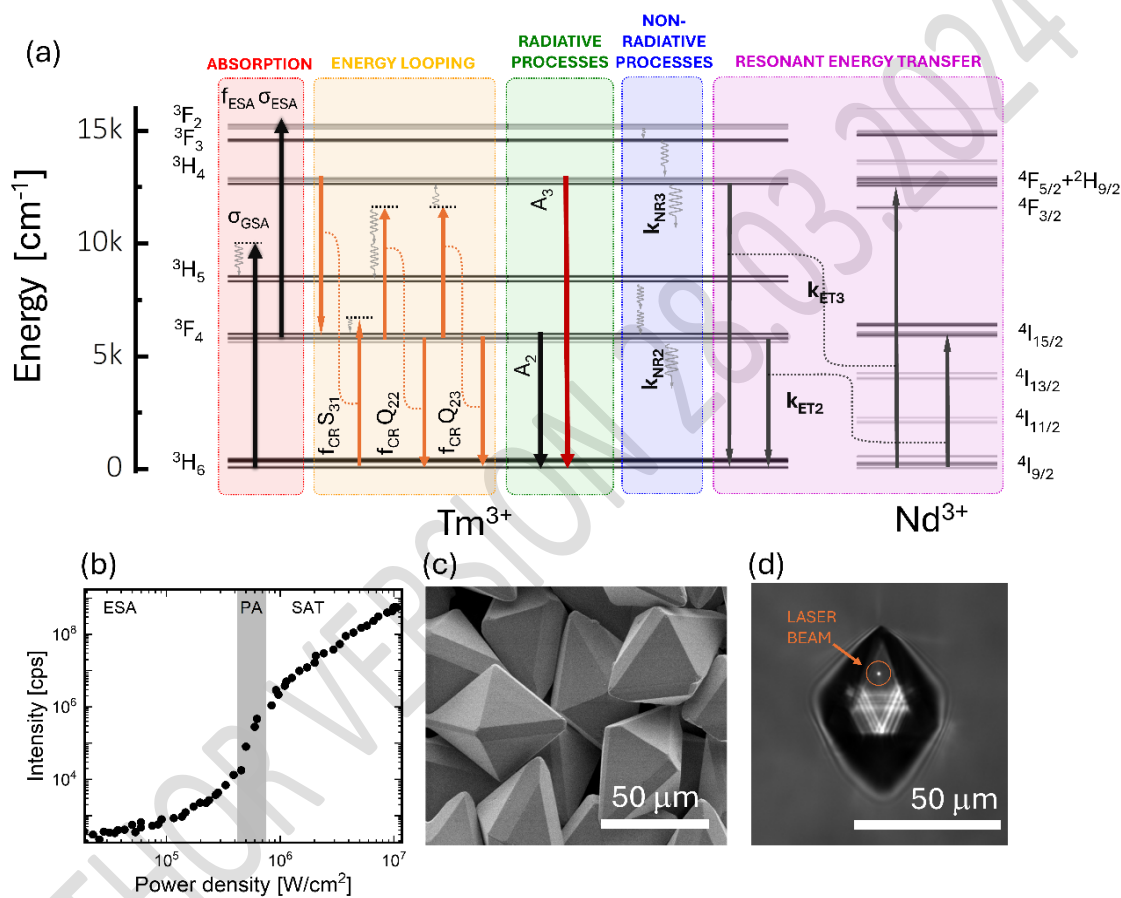

*Figure 2 (a) Energy level diagram of $Tm^{3+}$ showing radiative and non-radiative process leading to and accompanying PA process and energy diagram of the $Nd^{3+}$ acting as an acceptor and leading to parasite energy transfer from the $Tm^{3+}$ looping (mechanism I) and emitting (mechanism II) states; (b) experimental 800 nm pump power dependent emission showing three PA zones, i.e. low emission intensity ESA based up-conversion (ESA), PA emission (PA) and the saturation regime (SAT); (c) SEM image of $LiYF_4$:3%$Tm^{3+}$ microcrystals. (d) white field white light image, with 1059 nm laser beam focused on surface of the microcrystal.*

## 2. Results and discussion

Energy transfer in lanthanide ions depends largely on the spectral overlap integral between the respective levels and the physical distance between the donor and acceptor species. With the matching energy levels of the $Nd^{3+}$ and $Tm^{3+}$ ions, one can anticipate a relatively effective energy transfer between them despite the forbidden *f-f* dipole-dipole transitions. Here, we investigated the impact of presence and concentration of $Nd^{3+}$ ions on the pump-power-dependent PA emission intensity of $Tm^{3+}$ ions under 1059 nm photoexcitation. It is important to state, as discussed in SI, that ≤ 1% $Nd^{3+}$ ions doped in $LiYF_4$ crystals remain transparent under 1059 nm excitation at RT and does not exhibit PA-like behavior by itself, supporting the idea to use $Nd^{3+}$ ions as energy acceptors (through mechanism $Q_{III}$) from avalanching $Tm^{3+}$. The measured up-conversion spectra of the singly $Tm^{3+}$-doped sample reveal six emissions bands at 450 nm, 477 nm, 511 nm, 648 nm, 700 nm and the most intensely pronounced one at 800 nm, which can be ascribed to the $^1D_2 \rightarrow {^3F_4}$, $^1G_4 \rightarrow {^3H_3}$, $^1D_2 \rightarrow {^3H_5}$, $^1G_4 \rightarrow {^3H_6}$, $^3F_3 \rightarrow {^3H_3}$, and $^3H_4 \rightarrow {^3H_6}$ transitions, respectively (**Figure S4**). The introduction of $Nd^{3+}$ ions notably hindered the intensity of every emission band, which indicates efficient ET from $Tm^{3+}$ to $Nd^{3+}$ ions. Following ET pathways between donor and acceptor presented in **Table S1,** this can be either $(Tm;Nd)(^3H_4;{^4I_{9/2}}) \rightarrow (^3H_6; {^4F_{5/2}+^2H_{9/2}})$ or $(Tm;Nd)(^3F_4;{^4I_{9/2}}) \rightarrow (^3H_6; {^4I_{11/2}})$ $Tm^{3+}$ to $Nd^{3+}$ resonant energy transfers corresponding to $Q_I$ (LL) or $Q_{II}$ (EL) mechanisms, respectively. Co-doping the $Tm^{3+}$ PA system with $Nd^{3+}$ ions brings about a notable and gradual (as $Nd^{3+}$ concentration rises) alteration in the characteristic pump-power dependences (**Figure 3a, S4**). This co-doping hinders the PA process by reducing the nonlinearity (**Figure 3b**) and emission intensity (**Figure 3e**), as well as increasing the photoexcitation power threshold, $I_{th}$ (**Figure 3c**), necessary to achieve the PA regime. Excitation threshold $I_{th}$ is known to be proportional to the cross-section ESA coefficient and inversely proportional to the luminescence lifetime of the looping state and cross-section GSA absorption coefficient according to the $I_{th} \sim [\sigma_{ESA}/(\tau_2 \sigma_{GSA})]^{1/2}$.[29] Since $\sigma_{ESA}$ and $\sigma_{GSA}$ strictly depends on the wavelength choice and type of the absorbing ion, they remain constant for $Tm^{3+}$ ions independently of $Nd^{3+}$ co-doping. Hence, the increase in $I_{th}$ must be associated with a decrease in $\tau_2$, indicating the emergence of ET from the looping level. Moreover, the presence of acceptor ions reduced the maximum nonlinearity order, $S_{max}$ (**Figure 3b**). This suggests that the $Nd^{3+}$ ions hinder the smooth occurrence of the CR process, which is responsible for the nonlinear behavior observed in the PA emission.[29] To underscore the scale of quenching in the PA emission intensity, we collected Stokes 800 nm emission spectra (**Figure 3d, S6**) for comparative analysis (**Figure 3e**). At a constant excitation power of 4 MW/cm², it is evident that the introduction of 1% $Nd^{3+}$ doping results in a remarkable four-order magnitude reduction in the 800 nm PA emission intensity compared to the singly $Tm^{3+}$ doped sample, whereas the corresponding integrated Stokes emission is approximately four times less intense in the same and corresponding microcrystals (**Figure 3e**). The observed exceptional PA emission quenching in response to increasing acceptor concentration indicates that the PA phenomenon has huge application potential as a luminescent reporting feature and is suitable for the ultrasensitive sensing of physical (e.g., temperature, force, and plasmonic interactions), chemical (e.g., pH), and biological (FRET sensing) processes.

To quantify the superiority of PA emission over Stokes emission in acceptor sensing, we calculated the relative sensitivity ($S_R$) for both of them, as defined by $S_R = 1/I_{LUM} \cdot \Delta I_{LUM}/\Delta Conc_{Nd}$ (**Figure 3f**), which quantitatively depict the susceptibility of luminescence intensity to a unit change of acceptor concentration. PA emission at two fixed power densities (0.7 and 4 MW/cm²) demonstrated significantly higher $S_R$, reaching up to 900% at low concentrations and decreasing with higher concentrations down to 7-160% range. In contrast, Stokes emission covers a sensitivity range of lower but still reasonable 280 down to 60%. Notably, PA quenching is particularly suitable for detecting low acceptor concentrations owing to its superior sensitivity within this range. Nevertheless, such extreme sensitivities originating from the susceptibility of the PA mechanism to minute amounts of acceptor/quenching species have the other side of the coin, as PA may non-specifically be affected by other material factors (e.g., nano/micro crystal size, doping homogeneity, surface effects) (**Figure S7**), excitation (e.g., pump power density, pump power, and wavelength short- and long-term stability), or the properties of the actual sample (e.g., variation in pump power due to variations in light scattering). Therefore, it is critical to devise methodologies capable of reducing artifacts in order to enable reliable and reproducible sensing. Although one such approach has been proposed previously,[25] which suggests a comparison of luminescence under photoexcitation of avalanche materials through ESA and GSA, the lack of exact knowledge of the pumping intensity within scattering samples remains an open problem.

Another prominent evidence of the scale of quenching is the shortening of the Stokes luminescence lifetime. Stokes emission decays at 800 nm emission were measured and fitted with mono-exponential decay function $y = A_1 e^{-x/\tau}$ yielding time decay $\tau$ characteristic for the $^3H_4$ state (**Figure 3g**). ET efficiency can be determined based on the Stokes lifetime of the donor emission in presence ($\tau_{DA}$) or the absence of acceptor ($\tau_D$) in accordance with $\eta=(1-\tau_{DA}/\tau_D)\cdot 100\%$ formula. Thus, when considering $Nd^{3+}$ concentrations of 0.1, 0.2, 0.5, and 1%, the ET efficiency of the $^3H_4 \rightarrow {}^3H_6$ transition reaches 25, 39, 60 and 68%, respectively. Relative sensitivity based on the Stokes $^3H_4 \rightarrow {}^3H_6$ emission decay was comparable to the emission intensity variation approach (**Figure 3i**), and demonstrated relatively low sensitivity (i.e., 0.5 to 1%/1%$Nd^{3+}$) for 0.1-1% $Nd^{3+}$ concentration range. Measuring luminescence kinetics in the PA regime would be very informative, but because the looping level energetically falls within the NIR range (1670 nm), a specialized and typically noisy NIR detector is required to measure the emission and emission decays. An alternative method using the pump-probe technique was proposed to probe the looping level lifetime through 800 nm PA emission (**Figure S8**). Two subsequent pulses (pulse width d=10 ms) were used with different delay times ($\delta$ =1–70 ms), where the first pulse is responsible for the initial population of the $^3F_4$ level through GSA. The duration and power of the pulse are sufficient to initiate not only GSA and ESA but also the CR process leading to 800 nm emission. Because $^3F_4$ level is a reservoir level characterized by decay times of several milliseconds [30,31] and is responsible for preserving some oddment population that has been established with the first pulse, it is capable of facilitating faster and more efficient ESA during the subsequent pulse and thus should show enhanced luminescence intensity despite the pump pulses being equal in terms of energy and length. The ratio $I_2/I_1$ of the emission intensities induced by these two subsequent pulses ($I_2$) versus a single pulse ($I_1$), collected as a function of the time delay between the two pulses, provides information on how fast the $^3F_4$ decays (**Figure 3h**). When the ratio equals 1, there is no enhancement, therefore, during the delay time between the two pulses, the population of the $^3F_4$ level decays to zero. The $I_2/I_1$ ratio is called paired pulse facilitation (PPF = $I_2/I_1$) and was initially used to highlight the similarities between PA materials and biological synapses,[15] but here, its applicability in tracking the time decay of the looping state is demonstrated. The lifetimes of the $^3F_4$ level obtained from the exponential fitting of the PPF(d) dependence were investigated for $Nd^{3+}$ concentrations of 0% and 0.1% only, because the quenching effect of $Nd^{3+}$ ions is so strong that even at delay times ranging from 0.1 to 1 ms, higher concentrations did not demonstrate any emission intensity facilitation during the subsequent pulse. The lifetime decreases from 15.6 ms (in agreement with other literature results, i.e. 13.6 ms[30,31] for LiYF$_4$:3%Tm$^{3+}$) down to 5.2 ms upon co-doping with as little as 0.1% Nd$^{3+}$ (S$_R$=618%/1%Nd ; η = 62%), being a clear indication of the emergence of the new non-radiative process, which in PA mode is additionally competing with processes such as ESA and CR. These results prove that the population of $^3F_4$ looping-level growth is significantly hindered by low acceptor concentrations, resulting in a higher I$_{th}$.

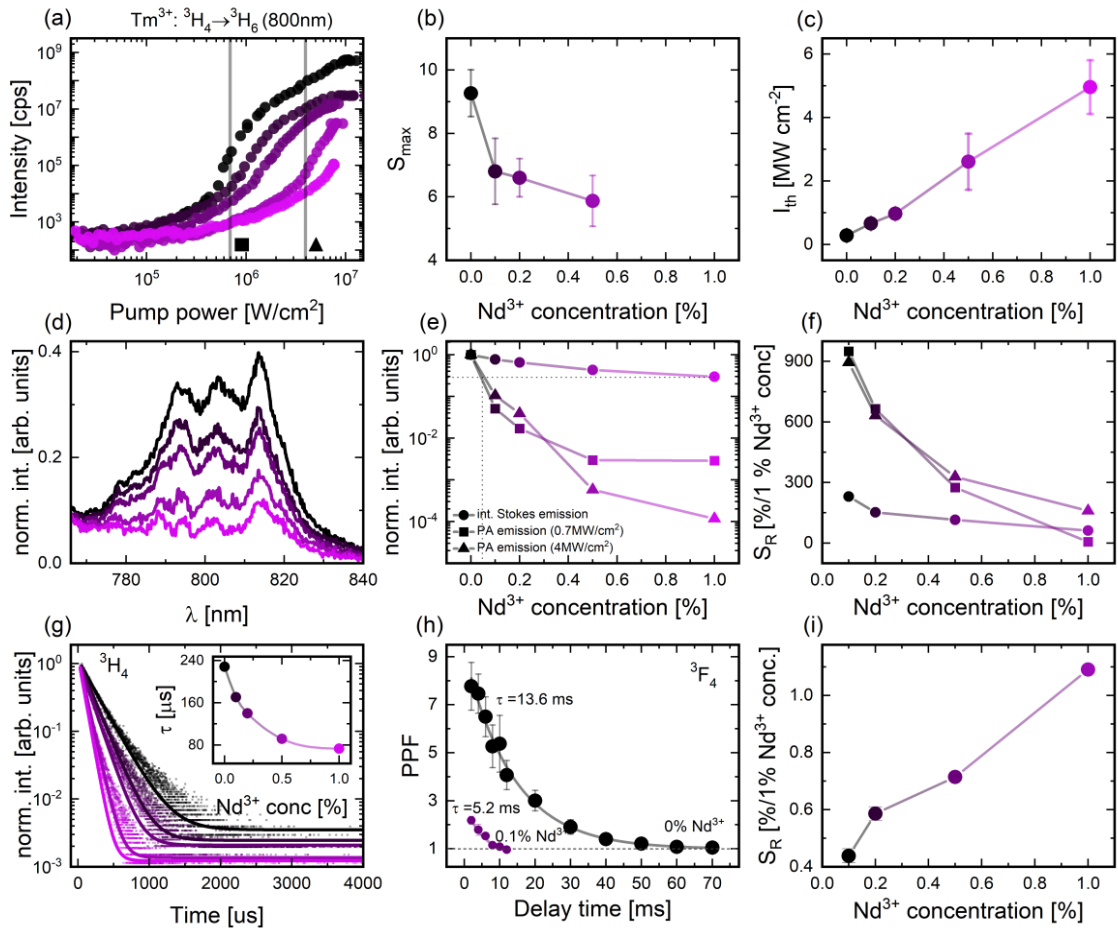

*Figure 3. Comparison of PA and Stokes emission quenching. (a) Pump power-dependent 800 nm ($^3H_4$ ® $^3H_6$) Tm$^{3+}$ PA emission upon quenching by co-doping 0.1, 0.2, 0.5 and 1% Nd$^{3+}$ acceptor; corresponding (b) averaged maximum order of nonlinearity and (c) power-excitation threshold in function of Nd$^{3+}$ concentration; (d) 800 nm Stokes emission quenching for 0, 0.1, 0.2, 0.5 and 1% Nd$^{3+}$ concentration; (e) normalized Stokes emission under 356 nm excitation wavelength (●) and PA emission at 0.7 MW/cm$^2$ (■) and 4 MW/cm$^2$ (▲) enables emission intensity quenching comparison; (f) relative sensitivity based on the Stokes and PA emission intensity; (g) normalized Stokes 800 nm emission decay fitted with mono-exponential decay, the inset depicts time decay in function of Nd$^{3+}$ concentration; (h) Paired pulse facilitation described as ratio of I$_2$/I$_1$ of 800 nm PA emission in function of delay time for the 0% and 0.1% Nd$^{3+}$ concentration depicting the looping level population decay; (i) relative sensitivity of the studied PA material determined based on time decay of 800 nm Stokes emission.*

It has been demonstrated that the PA process becomes progressively disrupted in the presence of Nd$^{3+}$ ions, but it is still unclear which specific mechanism underlies the observed changes, namely, whether PA is affected specifically by ET occurring by mechanism Q$_I$ (from looping $^3F_4$) and/or mechanism Q$_{II}$ (from emitting $^3H_4$). Therefore, differential rate equation-based modelling of PA (PADRE) was used to investigate how ET from the looping and emitting levels affects PA performance (**Equation S1**). Instead of reproducing the complex and interdependent PA behavior of the Nd$^{3+}$ co-doped LiYF$_4$:Tm$^{3+}$ system, a simplified PADRE model was employed to describe the PA emission at 800 nm, in which only $^3H_6$ (ground), $^3F_4$ (looping), and $^3H_4$ levels (emissive) of Tm$^{3+}$ and $^4F_{5/2}$, $^4I_{15/2}$ and $^4I_{9/2}$ (ground) Nd$^{3+}$ levels were considered.[12] To quantify the interaction between the two ions, the PADRE model was extended with ET from Tm$^{3+}$ to Nd$^{3+}$ by adding $k_{ET2}$ and $k_{ET3}$ parameters, which denote the nonradiative resonant energy transfer rates from the $^3F_4$ and $^3H_4$ levels to the respective $^4I_{15/2}$ and $^4F_{5/2}$ Nd$^{3+}$ levels.

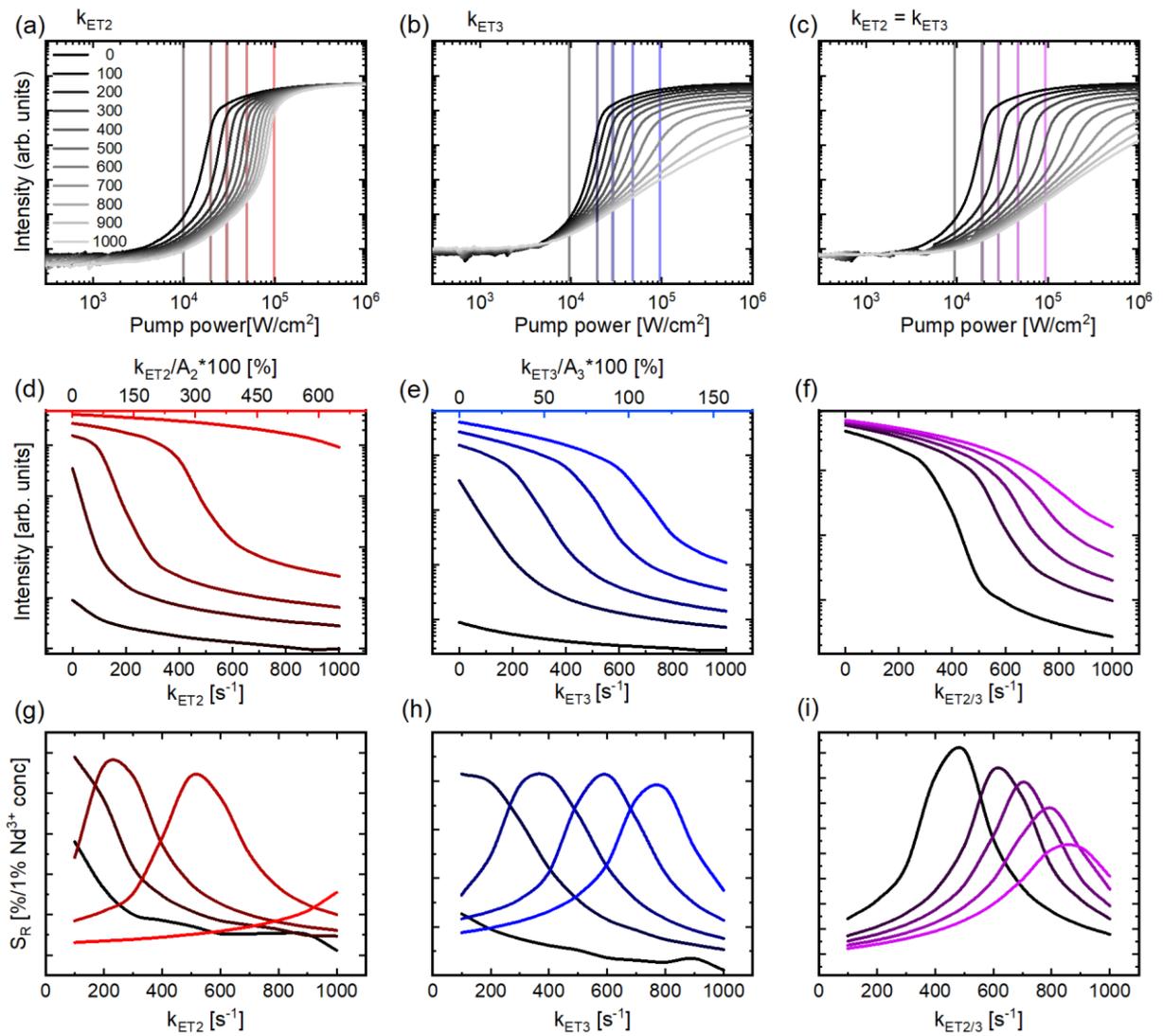

*Figure 4 Simulations of photon avalanche emission intensities in response to different quenching mechanisms and strenght. (a-c) The pump power-dependent 800 nm PA emission intensities were calculated based on the differential rate equations for varying energy*

*transfer rates (0-1000s$^{-1}$) and for scenarios involving ET from the looping state (a) via $k_{ET2}$, (b) from the emitting state via $k_{ET3}$, and (c) from both states equally with $k_{ET2}=k_{ET3}$; (d-e) the ET rate dependent PA emission intensites (from a-c correspondingly) at fixed power densities of 10, 20, 30, 50 and 100 kWcm$^2$ in fuction of (d) $k_{ET2}$, (e) $k_{ET3}$ and (f) $k_{ET2}=k_{ET3}$; the top axes of the d-e show the relative values as compared to $A_2$ and $A_3$ radiative rates (g-i) the relative, pump power dependent sensitivities derived from the determined PA emission intensities variation in fuction of (g) $k_{ET2}$, (h) $k_{ET3}$ and (i) $k_{ET2}=k_{ET3}$ respectively.*

Increasing the $k_{ET2}$ and $k_{ET3}$ factors in the range of 0-1000 s$^{-1}$ shows how individual ET from either looping, emitting, or both levels affects PA behavior (**Figure 4(a-c)**). At first glance, $k_{ET2}$ mainly affects the excitation threshold $I_{th}$, whereas $k_{ET3}$ significantly reduces the intensity and order of nonlinearity of the PA emission. To further investigate the influence of ET at each level on PA emission, the pump-power-dependent order of nonlinearities was determined (**Figure S9**). The factor $k_{ET2}$ not only increases $I_{th}$, but also contributes to the S factor. At the same time, $k_{ET3}$ does not seem to influence $I_{th}$ and is mainly responsible for the $S_{max}$ value decline. Since $k_{ET3}$ is a process directly competing with CR, it's clear that it affects substantially order of nonlinearity. Intriguingly, $k_{ET2}$ seems to have opposite effect on the order of nonlinearity to $k_{ET3}$ (**Figure S9**). This effect could be attributed to the fact that ET from the looping level contributes to the depopulation of the electrons excited through GSA, which are necessary to initialize PA, but at the PA regime are competing with CR. Power-excitation threshold $I_{th}$ is known to be inversely proportional to the luminescence lifetime of the looping state is solely affected by $k_{ET2}$ process. Further analysis of $S_{max}$ as a function of mechanisms $Q_I$ and $Q_{II}$ supports (**Figure S10-S11**) our hypothesis on the critical role of the looping-level population. The simulations and experimental studies are providing evidences for generic understanding of the quenching in PA materials and for further considerations of sensing of homogenous physical fields or quantities (e.g. temperature, pressure), when homogenous 3D mixture of donor and acceptor species is considered. But the next level of complexity, which is beyond the current work, should be expected when bio-specific interaction between avalanching nanoparticle (comprising thousands of interacting Tm$^{3+}$ ions) and the NP surface bound acceptors is taken into account. In such case, the volumetric interactions between looping Tm$^{3+}$ and homogenously incorporated acceptor species are replaced with a through the surface interactions of ANP (as an energy donor nanoparticle) and surface bound acceptors[32,33].

Similar to the experiments performed, the PADRE model-based simulated PA luminescence intensities (**Figure 4(d-f)**), as well as the theoretical $S_R$ were determined (**Figure 4(g-i)**) at fixed power densities (vertical lines in **Figure 4(a-c)**). The quenching-rate-dependent PA emission intensities monotonically and rapidly decreased in both the experiments (**Figure 3e**) and simulations (**Figure 4(d-f)**). In addition to understanding the fundamental interactions and susceptibility of PA emissions to the presence and characteristics of quenching strength and mechanism, these results clearly suggest the possibility of exploiting the studied mechanisms for multiple types of sensing. In metrology, the sensitivity and response dynamic range are key features of the transducing element, which are typically fixed. The simulation performed for PA luminescence suggests that the dynamic response range can be adjusted dynamically *in-situ* by adjusting the pump power intensities while preserving high sensitivity. Such a feature would be extremely valuable in situations where a wide detection range is required; however, experimental data do not equivocally confirm these observations and hypotheses. This inconsistency between the theoretical and experimental $S_R$ can be attributed to the limitations of the simplified three-level model of the donor, where the presence of an acceptor is expressed solely by two additional non-radiative quenching parameters ($k_{ET2}$ and $k_{ET3}$), which fails to fully replicate the dynamics occurring in the actual LiYF$_4$:3%Tm$^{3+}$ co-doped with Nd$^{3+}$ ion system. Moreover, the simulations offer a much wider and ideal dynamic range of the possible populations of ground and excited levels, whereas typical photodetectors are limited in this respect and are susceptible to background noise and other artifacts. Because, based on the simulation, both $k_{ET2}$ and $k_{ET3}$ contribute to $S_R$ range tunability, this effect should be further investigated with more bio-relevant acceptors with simple energetic structures, such as dyes, capable of absorbing either at 800 nm or 1670 nm.

## 3. Conclusions

PAs in nanomaterials have attracted increased interest owing to their extreme nonlinearity and the possibility of applying these interesting materials to numerous fields of science and technology. Although the advantages of the PA phenomenon (i.e., threshold behavior and extreme nonlinear response to photoexcitation, NIR to Vis/NIR conversion, etc.) and its target applications (i.e., labels for simple PASSI subdiffractional imaging and data storage, all-optical data reservoir processing, sensing of physical quantities, or chemical species) have been clearly identified, further progress may be hindered without a deep understanding of the susceptibility of PA to external chemical/physical factors, such as temperature, crystalline host, or quencher type and concentration. Although luminescence quenching in lanthanide-doped (nano)materials has been extensively studied, the unique character of the PA emission stemming from a complex combination of GSA/ESA and energy looping presents more challenges in understanding and managing newly developed materials and metrology with luminescent reporters. Therefore, the impact of energy transfer from photon-avalanching Tm$^{3+}$ ions, either from the looping or emitting states, or both, to intentionally admixed quenching Nd$^{3+}$ ions, was evaluated in LiYF$_4$ microcrystals. The experiments were planned to purposefully exclude surface effects that typically non-specifically quench the luminescence of conventional nanophosphors and to address specific resonant energy transfer to the acceptor, aiming to mimic the principles behind luminescence (bio)sensing. Interestingly, thulium emission quenching by Nd$^{3+}$ ions significantly differed depending on photoexcitation regime: a spectacular 4-order of magnitude PA emission quenching was observed in 1%Nd$^{3+}$ co-doped crystals (800 nm emission under 1059 nm photoexcitation), whereas only 4-fold Tm$^{3+}$ emission (800 nm emission under 475 nm photoexcitation) quenching was observed for the same 3%Tm 1%Nd LiYF$_4$ microcrystals under Stokes excitation as compared to pristine 3%Tm LiYF$_4$ microcrystals. Moreover, the experimental data evidenced not only significant PA emission intensity quenching, but also a maximum order of nonlinearity reduction and a linear pump power-excitation threshold increase. While no world records were attained in the non-linearity and further optimization of the materials are required for particular sensing applications, the studies allowed to understand generic mechanism behind PA quenching and the critical role of the looping level in the PA generation.

In addition to the experimental evaluation, a phenomenological model was developed to determine the nature and strength of this susceptibility to the presence and varying concentrations of quenching species. These modelling indicate that this is mostly the ET from the looping level, which leads to an increase in $I_{th}$ and $S_{max}$, whereas the ET from the emitting state should be associated with a decrease in $S_{max}$ and a decrease in the final emission intensity in the saturation regime. The extreme susceptibility of PA emission to quenching, compared to Stokes emission, makes the PA phenomenon and PA excitation scheme a novel ultrasensitive sensing mechanism with high and potentially *in-situ* tunable relative sensitivity. Moreover, both experiments and modelling confirmed the previous hypotheses that the extreme susceptibility of PA to quenching may potentially extend its effective (bio)detection range. Even though the resonant energy transfer occurs through Förster mechanism and is governed by distance power -6 ($r^{-6}$) law, it is important to note that 4-fold luminescence drop in response to quenching by 1%$Nd^{3+}$ in Stokes mode is equal to effects achieved in PA mode (dotted lines in **Figure 3e**) for as little as 0.08% $Nd^{3+}$. In the first approximation, such a comparison suggests that meaningfully smaller concentrations of acceptors may be detected in PA mode than in the conventional Stokes mode. This suggests that PA luminescence is equally quenched at D-A distance of 13.83 Å, as the Stokes luminescence with D-A distance of 6.25 Å – namely, at 2.2-fold longer range. Such effects are advantageous for (bio)chemical sensing, where the length of molecules responsible for specific biorecognition anchored to the surface of donor nanoparticles presents a challenge for ordinary up-converters owing to their relatively large size (typically 20 nm in diameter), lack of inherent biological specificity, and limited effective Förster distance from $Ln^{3+}$ donor ions to organic acceptors. Consequently, the FRET sensitivity is reduced, and most of the advantages of upconverting nanoparticles as luminescence donors for bioassays are dismissed.[32] Moreover, PA luminescence showed a very steep dynamic decline within the 0%-0.1% $Nd^{3+}$ concentration range, suggesting it is extremely and unproportionally more sensitive to tiny perturbations. Since PA process is far more complex than other up-conversion processes, one can deduce that it will surpass this problem by achieving higher effective Förster distance[34]. Having photon avalanching donor nanoparticle would thus solve a serious issue of conventional UCNPs based biosensors, preserving all other advantages the lanthanide ions and efficient anti-Stokes emission offer, such as narrowband absorption and emission, photo- and chemical stability and non-blinking as well as anti-Stokes NIR-to-NIR emission.

While qualitative correspondence exists between the experiments and modelling (e.g., **Figure 3e** vs. **Figure 4(d-f)**), the specific and detailed behaviors differ slightly. For example, the corresponding relative sensitivities $S_R$ based on the simulation were not fully consistent with those based on the experimental data; however, we expect the simulations to show a much broader concentration range than that of the experiments. The capability to *in-situ* tune sensitivity range – the conclusion which was derived from the modelling, is so far only weakly supported by the measurements (**Figure 3e-f**, ▲ vs ■, as compared to **Figure 4g-l**, respectively). Alternatively, we assume that some differences between the real Tm-Nd system and the simplified PADRE model occur; for example, the applied PADRE model may not include other factors, such as excited-state absorption by the $Nd^{3+}$ ions, energy cross relaxation between $Nd^{3+}$ ions (at increased $Nd^{3+}$ concentration), or back energy transfer from $Nd^{3+}$ to $Tm^{3+}$, which are currently excluded from the analysis.

Despite their impressively high relative sensitivity ($S_R$), PA-based sensors still need to address issues related to reproducibility and reliable readout. We assume that the extreme nonlinearity of the PA phenomenon, which defines the suitability of this mechanism for ultrasensitive detection, shows *the other side of the coin*. In the PA regime, the infinitesimal variation in the pumping intensity (of the light source, but potentially also due to light scattering in the heterogeneous biological samples) may not specifically modulate the output luminescence. This indicates that instabilities in the photoexcitation and photodetection paths are critical for the reproducibility of the readout. Therefore, special care must be taken to ensure the perfectly stable operation of the PA luminescence excitation and detection.

The advantages of PA emission, which were demonstrated and modelled in this work, indicate that highly nonlinear responsivity has key features that are of great value for the ultrasensitive detection and sensing of physical quantities, chemical environments, and biological reactions. The use of luminescence opens new possibilities for remote, nondestructive, nonelectric/magnetic field interference sensing with simple and robust optical instruments, even at subdiffraction spatial resolution[2]. Recent achievements in PA technology and an understanding of this phenomenon may therefore enable the design of photonic chips or complex platforms that combine sensing[8], data interpretation, [15] and storage[16].

## 4. Experimental

### Chemicals

The $LiYF_4$:TmNd microcrystals were prepared using commercially available reagents. Yttrium oxide $Y_2O_3$ (99.999%) and thulium oxide $Tm_2O_3$ (99.995%) were purchased from Chemicals 101 Corp. $Nd_2O_3$ (99.99%) and ammonium fluoride $NH_4F$ (≥98.0%) were purchased from Sigma Aldrich. Ethanol (96% pure p.a.) was purchased from POCH S.A. Lithium fluoride LiF and EDTA ≥99% (Ethylenediaminetetraacetic acid) were purchased from POL-AURA.

### Synthesis protocol

$LiYF_4$:$Tm^{3+}Nd^{3+}$ microcrystals were synthesized according to a previously published hydrothermal synthesis method, with some modifications.[35] First, 20 mmol of $RE(NO_3)_3$ ($Y(NO_3)_3$, $Tm(NO_3)_3$ and $Nd(NO_3)_3$) was dissolved in 15 mL of distilled water and added dropwise to 15 mL of a 1.06 g EDTA solution. The mixture was stirred vigorously at room temperature for 1 h. Simultaneously, 20 mmol LiF and 60 mmol $NH_4F$ were mixed with 30 mL distilled water. The RE-EDTA solution was then added dropwise to the fluoride solution under vigorous stirring. The thick white mixture was then stirred for 30 min and transferred to a 100 mL Teflon-lined autoclave. Synthesis was performed for 12 h at 200°C. The resulting microcrystals were separated from the supernatant and washed with distilled water and ethanol in an ultrasonic bath. The product was then dried on a hot plate at 100°C for 1 h.

### Experimental details

Pump-power-dependent PA measurements were conducted under a 1059 nm excitation source (single fiber laser diode) on a custom optical setup (**Figure S3**) consisting of an inverted microscope (Nikon Ti2 Eclipse), two photomultiplier tubes (PMT, PMT1001M, PMT2001M, Thorlabs), and a photon counter (quTAG, quTools). An arbitrary waveform generator (AWG Handyscope5, TiePie) controlling the laser diode voltage was used for the pump-probe experiments. The PA spectra were collected using a commercial Czerny-Turner spectrometer (Shamrock 500i). The Stokes spectra and emission decay were measured using a commercial Czerny-Turner photoluminescence spectrometer (FLS1000, Edinburgh Instruments). A more detailed description is provided in Supporting Information.

**Data modelling**

The simulations were executed using custom MATLAB code. PADRE equations were developed and studied in Refs. [12] and expanded here to account for the RET between Tm and Nd (Equation S1), were utilized to reproduce the nonlinear PA behavior and emission quenching.

**Author contributions**

Scientific concepts, ideas, and experimental designs were the result of interactions and discussions between the authors. M.Mi synthesized and characterized these particles. M.Ma performed all spectroscopic measurements, as well as modelling and analyses. A.B. conceived the idea, provided resources, and contributed to analyses and modelling. M.Ma and A. B. wrote the paper in coordination with all authors.

**Acknowledgement**

The authors declare that they have no competing interests. This research was funded in whole or in part by the following projects: 2018/31/B/ST5/01827 (AB) and 2021/43/B/ST5/01244 (AB, MMa, MMi), funded by the National Science Center, Poland. All data needed to evaluate the conclusions in the paper are present in the paper and/or the Supplementary Materials.